%-------------------------------------------------------------------
%    v7.tex   18 Nov 99    
%-------------------------------------------------------------------
%Variational study for Schroedinger's Hamiltonian
%  H=d^2/dx^2+Bx^2+A/x^2+lambda/x^alpha
%Joint paper Hall-Saad
% -------------------------------------------------------------------
\def\ptitle{\tiny Variational analysis for $\dots$}
% -------------------------------------------------------------------
\input psfig.sty
%-------------------------------------
%  generic unix fonts (lower case names)
% --------------------------------------------------------------------
\font\tr=cmr12                          % Our default
\font\bf=cmbx12                         % Redefinition
                         % Redefinition
\font\it=cmti12                         % Redefinition
\font\trbig=cmbx12 scaled 1500          % Main Title
                          % Theorems
\font\tiny=cmr10                        % Running title
% --------------------------------------------------------------------

\output={\shipout\vbox{\makeheadline
                                      \ifnum\the\pageno>1 {\hrule}  \fi
                                      {\pagebody}
                                      \makefootline}
                   \advancepageno}

\headline{\noindent {\ifnum\the\pageno>1
                                   {\tiny \ptitle\hfil
page~\the\pageno}\fi}}
\footline{}
% ---------------------------------------------------------------------
\tr
%--------------------------------------------------------------------
\def\bra{{\rm <}}    % bra ket:  math mode (to replace angle)
\def\ket{{\rm >}}    %   ket  >
%--------------------------------------------------------------------
% SPACING
% -------------------------------------------------------------------
\baselineskip 15 true pt  % draft 15  submit 18
\parskip=0pt plus 5pt
\parindent 0.25in
\hsize 6.0 true in
\hoffset 0.25 true in
% 6 in width with 1.25 in margins default = (6.5, 0)
\emergencystretch=0.6 in   % TEXBook p 107 : allows h-space
\vfuzz 0.4 in                            % page-length flexibility
\hfuzz  0.4 in                           % line-length flexibility
\vglue 0.1true in
\mathsurround=2pt                        % Default is 2pt
\topskip=24pt                            % Default is 10pt
% ---------------------------------------------------------------------
%  References
% ---------------------------------------------------------------------
\newcount\zz  \zz=0  % switch for printing references
\newcount\q   %  reference number
\newcount\qq    \qq=0  % starting reference number-1   (usually zero)

\def\pref #1#2#3#4#5{\frenchspacing \global \advance \q by 1  % paper reference
    \edef#1{\the\q}
       {\ifnum \zz=1 { %
         \item{[\the\q]}
         {#2} {\bf #3},{ #4.}{~#5}\medskip} \fi}}

\def\bref #1#2#3#4#5{\frenchspacing \global \advance \q by 1   % book reference
    \edef#1{\the\q}
    {\ifnum \zz=1 { %
       \item{[\the\q]}
       {#2}, {\it #3} {(#4).}{~#5}\medskip} \fi}}

\def\gref #1#2{\frenchspacing \global \advance \q by 1  % general reference
    \edef#1{\the\q}
    {\ifnum \zz=1 { %
       \item{[\the\q]}
       {#2}\medskip} \fi}}

 \def\sref #1{~[#1]}

\def\references#1{\zz=#1\parskip=2pt plus 1pt   % default is 0pt plus 1pt
   {\ifnum \zz=1 {\noindent \bf References \medskip} \fi} \q=\qq

\pref{\harr}{E. M. Harrell, Ann. Phys. }{105}{ 379 (1977)}{}
\pref{\klau}{J. R. Klauder, Acta Phys. Austriaca Suppl.}{11}{
341 (1973)}{}
\pref{\detw}{L. C. Detwiler and J. R. Klauder, Phys. Rev. D}{11}{1436
(1975)}{}
\pref{\simo}{B. Simon, J. Functional Anal.}{14}{295 (1973)}{}
\pref{\defa}{B. DeFacio and C. L. Hammer, J. Math. Phys.}{15}{1071
(1974)}{}

\pref{\agua}{V. C. Aguilera-Navarro, G.A. Est\'evez, and R.
Guardiola, J. Math. Phys.}{31}{99 (1990)}{}

\pref{\agub}{V. C. Aguilera-Navarro and R. Guardiola, J. Math.
Phys.}{32}{2135 (1991)}{}

\pref{\sol}{Solano-Torres, G. A. Est\'eves, F. M. Fern\'andez,
and G. C.
Groenenboom, J. Phys. A: Math. Gen.}{25}{3427 (1992)}{}

\pref{\flyn}{M. F. Flynn, R. Guardiola, and M. Znojil, Czech. J.
Phys.}{41}{1019 (1993)}{}

\pref{\znoa}{M. Znojil, Proc. Int. Conf. on Hadron Structure
}{91}{Proc. CSFR vol.1, 1 (1992)}{}

\pref{\znob}{M. Znojil, J. Math. Phys.}{34}{4914 (1993)}{}

\pref{\esta}{E. S. Est\'evez-Bret\'on and G. A.
Est\'evez-Bret\'on,  J. Math. Phys.}{34}{437 (1993)}{}

\pref{\hala}{R. Hall and N. Saad, Can. J. Phys.}{73}{493
(1995)}{}

\pref{\halb}{R. Hall and N. Saad, J. Phys. A: Math. Gen.}{31}{963
(1998)}{}

\pref{\halc}{R. Hall and N. Saad, J. Phys. A: Math. Gen.}{32}{133
(1999)}{}

\pref{\hald}{R. Hall, N. Saad and A. von Keviczky, J. Math.
Phys.}{39}{6345-51 (1998)}{}

\bref{\gold}{I. I. Gol'dman and D. V. Krivchenkov}{Problems
in Quantum mechanics}{Pergamon, London, 1961}{}

\bref{\slat}{L. J. Slater}{Confluent Hypergeometric
Functions}{At the University Press, Cambridge, 1960}{}

\pref{\bhat}{K. Bhattacharyya and R. K. Pathak, Int. J. Quantum
Chem.}{59}{219-226 (1996)}{}
\pref{\ball}{C. J. Ballhausen, Chem. Phys. Lett.}{146}{449 (1988)}{}

\pref{\sen}{P. Senn, Chem. Phys. letters}{154}{172 (1989)}{}

\bref{\pres}{H. W. Press, P.B. Flannery, A. S. Teukolsky and T.
W. Vetterling}{Numerical recipes in Pascal: The art of scientific
computing}{Cambridge University Press, Cambridge, 1989}{}

\pref{\diaz}{Carlos G. Diaz, Francisco M. Fern\'andez and
Eduardo A. Castro, J. Phys. A: Math. Gen.}{21}{L11 (1988)}{}

\pref{\papp}{E. Papp, Europhys. Lett.}{9}{309 (1989)}{}
\pref{\znoc}{M. Znojil, Phys. Lett. A}{255}{1 (1999)}{}
\pref{\znod}{M. Znojil, Phys. Lett. A}{259}{220 (1999)}{}

 } % end of ref list

 \references{0}    % Initialization of reference numbers
% -------------------------------------------------- end our ref.tex

%-------------------------------------------------------------------
%Title Page
%-------------------------------------------------------------------
\centerline{\bf\trbig Variational Analysis for a Generalized}
\centerline{\bf\trbig Spiked Harmonic Oscillator}
\medskip
\vskip 0.25 true in
\centerline{Richard L. Hall }
\medskip
{\leftskip=0pt plus 1fil
\rightskip=0pt plus 1fil\parfillskip=0pt\baselineskip 18 true pt
\obeylines
Department of Mathematics and Statistics, Concordia University,
1455 de Maisonneuve Boulevard West, Montr\'eal,
Qu\'ebec, Canada H3G 1M8.\par}
\vskip 0.25 true in
\centerline{Nasser Saad}
\medskip
{\leftskip=0pt plus 1fil
\rightskip=0pt plus 1fil\parfillskip=0pt\baselineskip 18 true pt
\obeylines
Department of Mathematics and Statistics,
Notre-Dame University(Louaize),
Zouk Mesbeh, Lebanon.\par}
\vskip 0.5 true in
%---------------------------------------------------------------------------

% Abstract
%---------------------------------------------------------------------------

\centerline{\bf Abstract}\medskip
A variational analysis is presented for the generalized spiked harmonic
oscillator Hamiltonian operator
$-{d^2\over dx^2}+Bx^2+{A\over x^2}+{\lambda\over x^{\alpha}}$,
where $\alpha$ is a real positive parameter. The formalism makes use of
a basis provided by exact solutions of Schr\"odinger's equation for the
Gol'dman and Krivchenkov Hamiltonian, and the corresponding matrix
elements that were previously found. For all the discrete eigenvalues
the method provides bounds which improve as the dimension $D$ of the
basis set is increased.  Extension to the $N$-dimensional case in arbitrary
angular-momentum subspaces is also presented. By minimizing over the free 
parameter $A,$ we are able to reduce substantially the number of basis functions 
needed for a given accuracy.
\vfill\eject
%---------------------------------------------------------------------------
\noindent{\bf I. Introduction}\medskip
%---------------------------------------------------------------------------
Since the fascinating work of Harrell\sref{\harr} on the ground-state
energy of the singular Hamiltonian
$H\equiv H_0+\lambda V=-{d^2/ {dx^2}}+x^2+{\lambda/ x^\alpha},\ x\in
[0,\infty],\ \alpha>0,$ known as the spiked harmonic oscillator Hamiltonian,
 the volume of research in this field has grown rapidly.  This is not only 
because of the
important applications of singular Hamiltonians to a wide variety of
problems in chemical, nuclear and particle physics, but also because of its
intrinsically interesting properties from the point view of mathematical
physics\sref{\klau-\defa}.  Most of these work\sref{\agua-\halc},
however, has focused on studying the spiked harmonic oscillator
Hamiltonian in one spatial dimension since
the interesting Klauder phenomenon\sref{\klau-\defa} associated with $H$
does not occur in higher dimensions.  Klauder\sref{\klau-\defa} has shown that, 
for
sufficiently singular potentials, $V$ cannot be turned off smoothly in the
Hamiltonian $H$ to restore the free Hamiltonian $H_0$.
Aguilera-Navarro et al\sref{\agua} employed variational
and perturbative schemes to solve the spiked harmonic oscillator problem for
the ground state energy. In their variational analysis of the
Hamiltonian $H$, Aguilera-Navarro et al employed the basis set of harmonic
oscillator eigenfunctions normalized in the interval $[0,\infty]$, i.e. the set 
of Hermite functions generated by the non-singular harmonic-oscillator
potential $x^{2}$.

Recently, we have obtained closed-form expressions\sref{\hald} for the
singular-potential integrals $<m|x^{-\alpha}|n>$ using the Gol'dman and 
Krivchenkov eigenfunctions\sref{\gold} for the singular Hamiltonian
$$
H_0=-{d^2\over {dx^2}}+Bx^2+{A\over x^2},\quad B>0, A\geq 0.\eqno(1.1)
$$
We present a variational analysis of the generalized spiked harmonic
oscillator Hamiltonian
$$
H\equiv H_0+\lambda V =-{d^2\over {dx^2}}+Bx^2+{A\over
x^2}+{\lambda\over
x^{\alpha}},\eqno(1.2)
$$
where $\lambda$ and $\alpha$ are positive parameters. To evaluate
the matrix elements of $x^{-\alpha}$, Hall et al\sref{\hald} used the
basis
set constructed with the normalized solutions of Schr\"odinger's
equation $H_0\psi=E\psi$, i.e.
$$
\left\{
\eqalign{
&\psi_n(x)\equiv |n\ket =C_nx^{{1\over 2}(1+\sqrt{1+4A})}e^{-{1\over
2}\sqrt{B} x^2}{}_1F_1(-n,1+{1\over 2}\sqrt{1+4A};\sqrt{B}x^2);\cr
&C_n^2=
{{2B^{{1\over 2}+{1\over 4}\sqrt{1+4A}}\Gamma(n+1+{1\over
2}\sqrt{1+4A})}
\over
{{n![\Gamma(1+{1\over 2}\sqrt{1+4A}})]^2}},\quad
n=0,1,2,\dots,\cr}\right.\eqno(1.3)
$$
where ${}_1F_1$ is the confluent hypergeometric function\sref{\slat}
$$
{}_1F_1(a,b;z)=\sum\limits_k {{(a)_kz^k}\over {(b)_kk!}},\
(a)_k=a(a+1)\dots(a+k-1)={{\Gamma(a+k)}\over {\Gamma(a)}}.
$$
Hall et al found, for $\alpha< 2\gamma$, that the matrix elements
$\bra m|x^{-\alpha}|n\ket$ are given by
$$
\eqalign{\bra m|x^{-\alpha}|n\ket&=(-1)^{n+m}B^{\alpha/4}
\sqrt{{\Gamma{(\gamma+m)}}\over {n!m!\ \Gamma{(\gamma+n)}}}\cr
&\times \sum_{k=0}^m (-1)^k{m \choose k}{{\Gamma(k+\gamma-{\alpha\over
2})\Gamma({\alpha\over 2}-k+n)}\over
{\Gamma(k+\gamma)\Gamma({\alpha\over
2}-k)}},\ \gamma=1+{1\over 2}\sqrt{1+4A},\ \cr}\eqno(1.4)
$$
in which each element has a factor which is a polynomial of degree $m+n$
in $\alpha$. Of particular interest are the matrix elements
$<0|x^{-\alpha}|n>$, which
can be obtained from Eq.(1.4), and read as follows:
$$
\bra 0|x^{-\alpha}|n \ket=(-1)^{n}B^{\alpha/4}
\sqrt{{\Gamma{(\gamma)}}\over {n!\
\Gamma{(\gamma+n)}}}{{\Gamma(\gamma-{\alpha\over 2})\Gamma({\alpha\over
2}+n)}\over {\Gamma(\gamma)\Gamma({\alpha\over 2})}},\
n=0,1,2,\dots.\eqno(1.5)
$$
The special case where $A=0$ and $B=1$ allows us to
recover the matrix elements provided by Aguilera-Navarro et
al\sref{\agua} for the operator $x^{-\alpha}$ in the harmonic oscillator 
representation, supplemented by the Dirichlet boundary condition $\psi(0)=0$.  
Indeed, substituting $A=0$ and $B=1$, and using the identities
$$
{{\Gamma(n+{3\over 2})}\over {\Gamma({3\over 2})}}={{(2n+1)!}\over
{2^{2n}n!}},\quad {{\Gamma(z+n)}\over {\Gamma({z})}}=(z+n-1)(z+n-2)\dots
(z+1)z,$$
after some algebraic simplification, we can easily show that Eq.(1.4) reduces to
Eq.(12) and Eq.(13), for $m=0$, of Ref.[\agua].  All
these expressions are valid for $\alpha<3$.

The purpose of this paper is to employ the variational method,
with the matrix elements (1.4), to solve the generalized singular
Schr\"odinger equation
$$\bigg[-{d^2\over {dx^2}}+Bx^2+{A\over x^2}+{\lambda\over
x^{\alpha}}\bigg]\psi=E\psi,\quad 0\leq x<\infty.\eqno(1.6)$$
The present
work is a generalization of the variational approach
of Aguilera-Navarro et al\sref{\agua} that used harmonic-oscillator
functions and
was restricted to ground state level ($B=1,A=0$).

The paper is organized as follows. In Sec. II, we outline the
variational method used to study (1.6), and we also extend the scope
to cover the $N-$dimension case. Some numerical results, and comparisons
with the results of Aguilera-Navarro et al, are presented in Sec. III. In
Sec. IV it is shown that a further optimization over the free parameter $A$ 
reduces substantially the number of basis functions needed to compute the 
eigenvalues to a given accuracy.\medskip
% ------------------------------------------------
\noindent{\bf II. The variational method}\medskip
% ------------------------------------------------
The first step in the variational method is to select a suitable
complete
set of basis functions that is adapted to the problem at hand. In the
variational analysis of the ground state energy of the singular
potential
$V(x)= x^2+{\lambda x^{-\alpha}},$
known in the literature as the spiked harmonic oscillator potential,
Aguilera-Navarro et al employed a basis set of harmonic oscillator
eigenfunctions normalized on the interval $[0,\infty]$ and vanishing at $x = 0$, 
i.e. the set of
odd Hermite functions generated by the non-singular harmonic-oscillator
potential $x^{2}$. A more effective basis set for the variational
analysis of such singular problems is the set of normalized wavefunctions (1.3)
because the singular characteristics of the potential are naturally built
into the wavefunctions.

Let $\psi(x)$ be a `trial function' for Hamiltonian $H$ given by (1.2),
and let us suppose that $\psi(x)$ can be expanded in terms of the basis
set $\psi_n(x)$ defined by (1.3).  Thus we have
$$
\psi(x)=\sum\limits_{n=0}^{D-1} a_n \psi_n(x).\eqno(2.1)
$$
The problem now is to minimize the eigenenergies of (1.2), with respect
to the variational parameters $a_n$, $n=0,1,\dots,D-1$, in the finite
dimensional subspace $H_{D}$ spanned by the $D$ functions
$\psi_0,\psi_1,\dots,\psi_{D-1}$.  This variational problem is
equivalent to diagonalizing the Hamailtonian (1.2) in the subspace $H_{D}.$
By increasing the dimension {\it D}, we can always improve
the results. Thus, we have to evaluate the matrix elements of the
Hamiltonian (1.2) in the basis (1.3). They can be separated into two
contributions
$$
H_{mn}=\bra m|H|n\ket\equiv \bra m|H_0|n \ket+\lambda\bra
m|x^{-\alpha}|n\ket,\quad
m,n=0,1,2,\dots,D-1.\eqno(2.2)
$$
Since $H_0$ is diagonal in the chosen basis, the first term on the
right hand side of (2.2) is the exact solutions of the Gol'dman and
Krivchenkov Hamiltonian, that is
$$
\bra m|H_0|n \ket =\sqrt{B}(4n+2+\sqrt{1+4A})\delta_{mn},\quad
m,n=0,1,2,\dots,D-1,\eqno(2.3)
$$
where $\delta_{mn}$ is the kronecker delta that equal to $1$ if $m=n$
and $0$ if m$\neq m$. The second term is
given by the matrix elements (1.4). Explicit expressions for the first
fifteen
matrix elements of $x^{-\alpha}$ are given here in the Appendix.

In order to extended the scope of this analysis to the $N$-dimensional 
Schr\"odinger
Equation (1.8), we observe first that the $A$ term has the dimensions of
kinetic energy, such as the term that appears in higher-dimensional
systems. We may therefore replace $A$ in Eq.(2.2) with
$$
A\rightarrow A + (l+{1\over 2}(N-1))(l+{1\over 2}(N-3)),\quad N\geq 2,\eqno(2.4)
$$
where the unperturbed energy levels (2.3) becomes in this case
$$
\bra m|H_0|n \ket =2\sqrt{B}\bigg(2n+1+\sqrt{A+(l+{N\over 2}-1)^2}\bigg)\delta_{mn},\quad N\geq 2.\eqno(2.5)
$$
Thus, Eq.(2.2) becomes
$$H_{mn}=2\sqrt{B}\bigg(2n+1+\sqrt{A+(l+{N\over 2}-1)^2}\bigg)\delta_{mn}+\lambda\bra
m|x^{-\alpha}|n\ket,\quad
m,n=0,1,2,\dots,D-1,\eqno(2.6)$$
where the matrix elements $\bra m|x^{-\alpha}|n\ket$ becomes
$$
\eqalign{\bra m|x^{-\alpha}|n\ket&=(-1)^{n+m}B^{\alpha\over 4}
\sqrt{{\Gamma{(m+1+\sqrt{A+(l+N/2-1)^2})}}\over {n!m!\ 
\Gamma{(n+1+\sqrt{A+(l+N/2-1)^2})}}}\cr
&\times \sum_{k=0}^m (-1)^k{m \choose k}{{\Gamma(k-{\alpha\over 2}+1+\sqrt{A+(l+N/2-1)^2}
)\Gamma({\alpha\over 2}-k+n)}\over {\Gamma(k+l+\sqrt{A+(l+N/2-1)^2})\Gamma({\alpha\over 2}-k)}},\quad N\geq 2.\cr}
\eqno(2.7)
$$
We have omitted the case $N=1$ because this curious singular problem in
one dimension has features\sref{\bhat-\sen} that are not in harmony with our
main purpose.

To recover the results of Aguilera-Navarro et al, we substitute $A=0$, $B=1$, $N=3$,and $l=0$ in Eq.(2.6). In other words, we
obtain a general variational expression (2.2) that treats the solution of 
Schr\"odinger's
equations (1.8), and also the work of Aguilera-Navarro et al in a
single formulation.
\medskip
% -----------------------------------------------------
\noindent{\bf III. Some numerical results}\medskip
% -----------------------------------------------------
From Eqs. (2.2), (2.3), and (1.4) and the results given in the Appendix,it can 
be readily seen that the first variational approximation (subspace
of dimension 1) to the ground state eigenvalues of the Hamiltonian (1.2)
is 
$$E_{0}=H_{00}=\bra 0|H|0\ket = 2\sqrt{B}\gamma+\lambda B^{\alpha\over
4}{{\Gamma(\gamma-{\alpha\over 2})}\over {\Gamma(\gamma)}},\
\gamma=1+{1\over 2}\sqrt{1+4A}.\eqno(3.1)$$
For the case $A=0$ and $B=1$, $E_{0}$ reduces to
$$E_{(0)}= 3+\lambda {{\Gamma({3-\alpha\over 2})}\over {\Gamma({3\over
2})}},\eqno(3.2)$$
which coincides with the ground-state energy expression for the
spiked harmonic oscillator obtained by Aguilera-Navarro et al, i.e.
Eq.(5.1) in Ref.[\agua].

When $D=2$ the diagonalization can also be performed analytically, by
means of the secular equation, and we obtain
$$\eqalign{&E_{\pm}=
{1\over 2}
\bigg[4\sqrt{B}(1+\gamma)+{\lambda\over 4} B^{\alpha\over
4}(\alpha^2-2\alpha+8\gamma){{\Gamma(\gamma-{\alpha\over 2})}\over
\Gamma(\gamma+1)}\pm \cr
&
\sqrt{
16B+2\lambda B^{\alpha+2\over 4}\alpha(\alpha-2)
{{\Gamma(\gamma-{\alpha\over 2})}\over \Gamma(\gamma+1)}+{\lambda^2\over
16}B^{\alpha\over
2}\alpha^2((\alpha-2)^2+16\gamma)\bigg[{{\Gamma(\gamma-{\alpha\over
2})}\over \Gamma(\gamma+1)}\bigg]^2}\bigg],\cr}\eqno(3.3)
$$
where $E_{0}=E_{-}$ and $E_{1}=E_{+}$. Again, if we put $A=0,$ $B=1,$
and $\alpha=5/2$ we obtain
$$E_{\pm}=
{1\over 2}
\bigg[10+{53\over 24}\lambda {{\Gamma({3-\alpha\over 2})}\over
\Gamma({3\over 2})}\pm
\sqrt{
16+{5\over 2}\lambda
{{\Gamma({3-\alpha\over 2})}\over \Gamma({3\over 2})}+{2425\over
576}\lambda^2\bigg[{{\Gamma({3-\alpha\over 2})}\over \Gamma({3\over
2})}\bigg]^2}\bigg],\eqno(3.4)
$$
which coincides with Eq.(5.2) in Ref.[6].

For higher values of the variational space dimension $D$ we have to use a 
numerical diagonalization procedure such as that of Jacobi\sref{\pres} to find 
the eigenvalues of the matrix $\cal{H}$ given by
$$\cal{H} = \pmatrix{
  H_{00}&H_{01}&\dots&H_{0D-1}\cr
  H_{10}&H_{11}&\dots&H_{1D-1}\cr
  \dots&\dots&\dots&\dots\cr
            H_{D-10}&H_{D-11}&\dots&H_{D-1D-1}\cr}\eqno(3.5)
$$
%%\rightarrow \rm{diagonalization\ procedure}\rightarrow E_n\leq
%%\rm{diag.}(E_0,E_1,...,E_{D-1})\eqno(3.5)$$
Table I  shows the convergence of these results, when $\alpha=2.5$, for the 
ground state energy (i.e. $A=0, B=1$) of the spiked harmonic oscillator, and 
selected values of $\lambda$. For this special case with $A = 0$ and 
$\alpha=5/2$, our results confirm those of Ref.[\agua] ({\it their} Table I), 
except for minor rounding errors: it appears that the results of 
Aguilera-Navarro et al have been truncated rather than rounded. It is important 
to mention here that, although the matrix element
$x_{33}^{-\alpha}$, reported in Ref.[\agua], contains some errors in the
coefficients for the exponent $\alpha$, the results in Table I of Ref.[\agua]
are correct. Indeed, $x_{33}^{-\alpha}$ should read:
$$x_{33}^{-\alpha}={{\Gamma({{3-\alpha}\over 2})}\over {7!\Gamma({3\over
2})}}(\alpha^6-6\alpha^5+106\alpha^4-384\alpha^3+2080\alpha^2-3408\alpha+5040)
\eqno(3.6)$$
instead of 
$$x_{33}^{-\alpha}={{\Gamma({{3-\alpha}\over 2})}\over {7!\Gamma({3\over
2})}}(\alpha^6-6\alpha^5+106\alpha^4-454\alpha^3+1660\alpha^2-3968\alpha+5040)
\eqno(3.7)$$
as quoted by Aguilera-Navarro et al\sref{\agua}.  The results quoted in
their Table I must therefore have been calculated with the correct formula.
We have also corrected in the Appendix here some errors in the coeffecients of the general matrix element $x^{-\alpha}_{33}$ presented in Ref.\sref{\hald}.

In Table (II) here we report the upper bounds $E_{nl}=E_{00}^N$ obtained
for the Hamiltonian $H=-{{d^2}\over {dx^2}}+x^2+{10\over x^{1.9}}$ in
spatial dimensions $N = 2$ to $N=10$. Now the value of $A$ depends on the 
angular momentum $\ell$ and the number of spatial dimension $N$ as determined by 
Eq.(2.4).  As we mentioned above, the method we have discussed in Sec.II 
provides bounds on {\it all} the eigenvalues in a
given angular momentum subspace. We present in Table (III) our results for the
eigenvalues $E_{nl}=E_{21}^N$ for spatial dimensions $N = 2$ to $N=10;$
the potential and the corresponding wave functions are shown in Fig.(1).
For comparison, we have integrated Schr\"odinger's equation numerically,
with the Hamiltonian given by (1.2), for different values of exponent
$\alpha$. To avoid difficulties caused by the singular character of the
potential near the origin, we begin the integrations near the potential
minimum and integrate in both directions, away from the starting point.  A
similar approach has been described by Diaz et al\sref{\diaz}.\medskip
%kludge
%------------------------------------------------------
\noindent{\bf IV. A further variational refinement}\medskip
%------------------------------------------------------
We now introduce another variational adjustment that will substantially
reduce the number of the basis elements required to compute the eigenvalues
of the Hamiltonian, i.e.
$$
H=-{{d^2}\over {dx^2}}+{{(l+{1\over 2}(N-1))(l+{1\over 2}(N-3))}\over
x^2}+Bx^2+{\lambda\over x^{\alpha}}. \eqno(4.1)
$$
We notice first that the Hamiltonian (4.1) can be written as
$$
H=-{{d^2}\over {dx^2}}+{{(l+{1\over 2}(N-1))(l+{1\over 2}(N-3))}\over
x^2}+Bx^2+{A\over x^2}+\bigg({\lambda\over x^{\alpha}}-{A\over
x^2}\bigg)\eqno(4.2)
$$
where $A$ is additional variational parameter, different from zero, to
be determined later. In this case Eq.(2.2) becomes
$$
H_{mn}=2\sqrt{B}(2n+1+\sqrt{A+(l+{N/2}-1)^2})\delta_{mn}+\lambda\bra
m|x^{-\alpha}|n\ket-A\bra m|x^{-2}|n\ket,\eqno(4.3)
$$
where $\bra m|x^{-\alpha}|n\ket$ is given by (2.7) and $\bra m|x^{-2}|n\ket$ is obtained by setting $\alpha=2$ in this formula. Upper bounds to the eigenvalues $E_{nl}^N$ are again provided by finding the 
eigenvalues of the matrix $\cal{H}$ given by Eq.(3.5), but with the entries 
depending on $A$ according to (4.3).

For example, when the variational space has dimension $D = 1,$ the lowest 
eigenvalue of the problem in $N$-dimensions labelled by $l$ is determined by 
$$H_{00}=\sqrt{B}(2\gamma-{A\over \gamma-1})+\lambda
B^{\alpha/4}{{\Gamma(\gamma-{\alpha\over 2}})
\over \Gamma(\gamma)},\eqno(4.4)
$$
where $\gamma=1+\sqrt{A+(l+{N\over 2}-1)^2}$. For $N=3,l=0,$ the minimum of 
$H_{00}$ with respect to $A$ for $\alpha=2$ occurs at $A=\lambda$, thus yielding 
the exact solution of the spiked harmonic oscillator
$E_0=\sqrt{B}(1+\sqrt{1+4A})$. In fact, a `good' general estimate for the value 
of $A$ is $A=\lambda.$  This rough estimate for $A$ reduces substantially the 
number
of the basis function needed to compute the eigenvalues;  by minimizing over 
$A$, we obtain even better upper bounds, as Table (IV) clearly indicates.
\medskip
% -----------------------------------------------------
\noindent{\bf V. Conclusion}\medskip
% -----------------------------------------------------+
We have generalized the work of Aguilera-Navarro et al to
treat the more general spiked harmonic oscillator problem (1.2). In
particular, we have presented a variation method to solve the
interesting $N$-dimensional spiked harmonic oscillator problem with
the $x^{-\alpha}$ singular term.  The present work is limited by the
necessary condition $\alpha < 3.$  Some interesting results for $\alpha \geq 3$
may be found, for example, in Refs.[\papp - \znod]. 

We hope that our work will encourage further research into the spectra
generated by this interesting class of singular potentials. It is very
clear from our results that use of a variational basis which is itself
derived from a related soluble singular problem leads to very effective
approximation methods for more general problems of this singular type. The 
presence of the free parameter $A$ in the class of soluble problems allows a 
further refinement in the upper energy estimates.
\bigskip
\noindent{\bf Acknowledgment}
\medskip Partial financial support of this work under Grant No. GP3438
from the Natural Sciences and Engineering Research Council of Canada is
gratefully acknowledged.
\bigskip

\vfill\eject
% ------------------------------------------
\references{1}
% ------------------------------------------
\vfil\eject
% -------------------------------------------------------------------------

\noindent{\bf Appendix: Some explicit forms of the matrix elements
$\bra m|x^{-\alpha}|n\ket$}\medskip
% -------------------------------------------------------------------------

\noindent We present the first fifteen matrix elements of $x^{-\alpha}$
that we  used to compute the variational eigenvalues in the next
sections. Some errors in the coefficients of $x_{33}^{-\alpha}$ in Ref.[16]
have been corrected. In terms of the parameter $\gamma=1+{1\over 2}\sqrt{1+4A}$, the
explicit matrix elements, from Eq.(1.4) and Eq.(1.5), are as follows:

\noindent $x_{nm}^{-\alpha}\equiv \bra m|x^{-\alpha}|n\ket$
\medskip
\noindent {$x_{00}^{-\alpha}=$}$B^{\alpha\over 4}{{\Gamma(-{\alpha\over
2}+\gamma)}\over {\Gamma(\gamma)}}$
\medskip
\noindent {$x_{01}^{-\alpha}=$}$-
B^{\alpha\over 4}
{\alpha\over 2\sqrt{\gamma}}
{{\Gamma(-{\alpha\over 2}+\gamma)}\over \Gamma(\gamma)}$
\medskip
\noindent {$x_{02}^{-\alpha}=$}$
B^{\alpha\over 4}
{{\alpha(\alpha+2)}\over {2^2\sqrt{2!\gamma(\gamma+1)}}}
{{\Gamma(-{\alpha\over 2}+\gamma)}\over \Gamma(\gamma)}$
\medskip
\noindent {$x_{03}^{-\alpha}=$}$-
B^{\alpha\over 4}
{{\alpha(\alpha+2)(\alpha+4)}\over
{2^3\sqrt{3!\gamma(\gamma+1)(\gamma+2)}}}
{{\Gamma(-{\alpha\over 2}+\gamma)}\over \Gamma(\gamma)}$
\medskip
\noindent {$x_{04}^{-\alpha}=$}$
B^{\alpha\over 4}
{{\alpha(\alpha+2)(\alpha+4)(\alpha+6)}\over
{2^4\sqrt{4!\gamma(\gamma+1)(\gamma+2)(\gamma+3)}}}
{{\Gamma(-{\alpha\over 2}+\gamma)}\over \Gamma(\gamma)}$
\medskip
\noindent {$x_{11}^{-\alpha}=$}${{B^{\alpha\over 4}}}
{{(\alpha^2-2\alpha+4\gamma)}\over 2^2}
{{\Gamma(-{\alpha\over
2}+\gamma)}\over {\Gamma(\gamma+1)}}$
\medskip
\noindent {$x_{12}^{-\alpha}=$}${{-B^{\alpha\over 4}}}
{{\alpha(\alpha^2-2\alpha+8\gamma)}\over {2^3\sqrt{2!(\gamma+1)}}}
{{\Gamma(-{\alpha\over
2}+\gamma)}\over {\Gamma(\gamma+1)}}$
\medskip
\noindent {$x_{13}^{-\alpha}=$}${{B^{\alpha\over 4}}}
{{\alpha(\alpha+2)(\alpha^2-2\alpha+12\gamma)}\over
{2^4\sqrt{3!(\gamma+1)(\gamma+2)}}}
{{\Gamma(-{\alpha\over
2}+\gamma)}\over {\Gamma(\gamma+1)}}$
\medskip
\noindent {$x_{14}^{-\alpha}=$}${{-B^{\alpha\over 4}}}
{{\alpha(\alpha+2)(\alpha+4)(\alpha^2-2\alpha+16\gamma)}\over
{2^5\sqrt{4!(\gamma+1)(\gamma+2)(\gamma+3)}}}
{{\Gamma(-{\alpha\over
2}+\gamma)}\over {\Gamma(\gamma+1)}}$
\medskip
\noindent {$x_{22}^{-\alpha}=$}${{B^{\alpha\over 4}}}
{{\alpha^4-4\alpha^3+(12+16\gamma)\alpha^2-(16+32\gamma)\alpha+32\gamma(1+\gamma
)}\over
{2^4\sqrt{2!2!}}}
{{\Gamma(-{\alpha\over
2}+\gamma)}\over {\Gamma(\gamma+2)}}$
\medskip
\noindent {$x_{23}^{-\alpha}=$}
${{-B^{\alpha\over 4}}}
{{\alpha(\alpha^4-4\alpha^3+(20+24\gamma)\alpha^2-(32+48
\gamma)\alpha+96\gamma(1+\gamma))}\over{2^5\sqrt{2!3!(\gamma+2)}}}
{{\Gamma(-{\alpha\over
2}+\gamma)}\over {\Gamma(\gamma+2)}}$
\medskip
\noindent {$x_{24}^{-\alpha}=$}
${{B^{\alpha\over 4}}}
{{\alpha(\alpha^5-2\alpha^4+(20+32\gamma)\alpha^3+8\alpha^2+(-96+64\gamma+192
\gamma^2)\alpha+384\gamma(\gamma+1)
)}\over{2^6\sqrt{2!4!(\gamma+2)(\gamma+3)}}}
{{\Gamma(-{\alpha\over
2}+\gamma)}\over {\Gamma(\gamma+2)}}$
\medskip
\noindent $x_{33}^{-\alpha}=$ ${{{B^{\alpha\over 4}}}\over
{2^6\sqrt{3!3!}}}
\bigg(\alpha^6-6\alpha^5+(52+36\gamma)\alpha^4-(168+144\gamma)\alpha^3+(352+720
\gamma+288\gamma^2)
\alpha^2-(384+1152\gamma+576\gamma^2)\alpha+384\gamma(1+\gamma)(2+\gamma)\bigg)
{{\Gamma(-{\alpha\over 2}+\gamma)}\over {\Gamma(\gamma+3)}}$
\medskip
\noindent $x_{34}^{-\alpha}=$ -${{{B^{\alpha\over 4}}}\over
{2^7\sqrt{3!4!(\gamma+3)}}}
\bigg(\alpha^7-6\alpha^6+(76+48\gamma)\alpha^5-(264+192\gamma)\alpha^4+(832+1536
\gamma+576\gamma^2)
\alpha^3-
(1152+2688\gamma+1152\gamma^2)\alpha^2+
1536\gamma(\gamma+1)(\gamma+2)\alpha\bigg)
{{\Gamma(-{\alpha\over 2}+\gamma)}\over {\Gamma(\gamma+3)}}$
\medskip
\noindent $x_{44}^{-\alpha}=$ ${{{B^{\alpha\over 4}}}\over
{2^8\sqrt{4!4!}}}
\bigg(\alpha^8-8\alpha^7+(136+64\gamma)\alpha^6-(704+384\gamma)\alpha^5+(3856+44
80\gamma+1152\gamma^2)
\alpha^4-(10880+15360\gamma+4608\gamma^2)\alpha^3+$
$(19200+48640\gamma+32256\gamma^2+6144\gamma^3)\alpha^2-
(18432+67584\gamma+55296\gamma^2+12288\gamma^3)\alpha
+6144\gamma(\gamma+1)(\gamma+2)(\gamma+3)\bigg)
{{\Gamma(-{\alpha\over 2}+\gamma)}\over {\Gamma(\gamma+4)}}$
\bigskip
\vfill\eject

% -----------------------------------------------------------------------------

{\noindent {\bf Table(I)}~~~The ground-state eigenvalue of Schr\"odinger's 
equation $H\psi=E\psi$, where $A = 0$ and $H=-{{d^2}\over {dx^2}}+x^2+{\lambda\over x^{2.5}}$,
obtained by diagonalization of the $D\times D$ matrix elements with $D=1,2,10,20,30$. The ``exact'' values $E$ were obtained by direct numerical integration of Schr\"odinger's
equation.
\bigskip
\centerline{
\vbox{\tabskip=0pt\offinterlineskip
\def\tablerule{\noalign{\hrule}}
\def\vr{\vrule height 12pt}
\halign to360pt{\strut#\vr&#
\tabskip=0em plus1em
&\hfil#\hfil&\vrule#
&\hfil#\hfil&\vrule#
&\hfil#\hfil&\vrule#
&\hfil#\hfil&\vrule#
&\hfil#\hfil&\vrule#
&\hfil#\hfil&\vrule#
&\hfil#\hfil&\vrule#\tabskip=0pt\cr
\tablerule&&$\lambda$&&$E^{(1)}$&&$E^{(2)}$&&$E^{(10)}$&&$E^{(20)}$&&$E^{(30)}$&
&$E$&
\cr\tablerule
&&$1000$&&$4094.062~692$&&$324.897~482$&&
$44.967~048$&&$44.955~485$&&$44.955~485$&&$44.955~485$&\cr\tablerule
&&$100$&&$412.106~269$&&$36.802~319$&&$17.541~891$&&$17.541~890$&&$17.541~890$&&
$17.541~890$
&\cr\tablerule
&&$10$&&$43.910~627$&&$7.951~034$&&$7.735~637$&&$7.735~135$&&$7.735~114$&&$7.735
~111
$&\cr\tablerule
&&$5$&&$23.455~313$&&$6.304~224$&&$6.297~319$&&$6.296~710$&&$6.296~566$&&$6.296~
473
$&\cr\tablerule
&&$1$&&$7.091~063$&&$4.688~098$&&$4.354~248$&&$4.329~430$&&$4.323~263$&&$4.317~3
12
$&\cr\tablerule
&&$0.5$&&$5.045~531$&&$4.216~200$&&$3.919~692$&&$3.882~149$&&$3.869~547$&&$3.848
~553
$&\cr\tablerule
&&$0.1$&&$3.409~106$&&$3.366~867$&&$3.316~061$&&$3.302~484$&&$3.296~024$&&$3.266
~871
$&\cr\tablerule
&&$0.05$&&$3.204~553$&&$3.193~800$&&$3.177~840$&&$3.172~753$&&$3.170~127$&&$3.15
2~420
$&\cr\tablerule
&&$0.01$&&$3.040~911$&&$3.040~476$&&$3.039~702$&&$3.039~409$&&$3.039~244$&&$3.03
6~665
$&\cr\tablerule
&&$0.005$&&$3.020~455$&&$3.020~346$&&$3.020~148$&&$3.020~071$&&$3.020~027$&&$3.0
19~086
$&\cr\tablerule
&&$0.001$&&$3.004~091$&&$3.004~087$&&$3.004~079$&&$3.004~075$&&$3.004~074$&&$3.0
04~014
$&\cr\tablerule}}
}}

\vfil\eject
% ---------------------------------------------------------------------
{\noindent {\bf Table(II)}~~~Upper bounds $E_{00}^N$ for $A = 0$ and $H=-{{d^2}\over {dx^2}}+x^2+{10\over x^{1.9}}$ for dimension $N = 2$ to $10$, obtained by diagonalization of the $D\times D$ matrix $\cal{H}$, $D=1,2,10,20,30$. The ``exact'' values $E$ were obtained by direct numerical integration of Schr\"odinger's equation.

\bigskip
\centerline{
\vbox{\tabskip=0pt\offinterlineskip
\def\tablerule{\noalign{\hrule}}
\def\vr{\vrule height 12pt}
\halign to350pt{\strut#\vr&#
\tabskip=0em plus1em
&\hfil#\hfil&\vrule#
&\hfil#\hfil&\vrule#
&\hfil#\hfil&\vrule#
&\hfil#\hfil&\vrule#
&\hfil#\hfil&\vrule#
&\hfil#\hfil&\vrule#
&\hfil#\hfil&\vrule#\tabskip=0pt\cr
\tablerule&&$N$&&$E^{(1)}$&&$E^{(2)}$&&$E^{(10)}$&&$E^{(20)}$&&$E^{(30)}$&&$E$&
\cr\tablerule
&&$2$&&$196.700~853$&&$9.092~284$&&$8.485~580$&&$8.485~399$&&$8.485~384$&&$8.485
~378
$&\cr\tablerule
&&$3$&&$21.236~010$&&$8.698~978$&&$8.564~442$&&$8.564~364$&&$8.564~358$&&$8.564~
356$&\cr\tablerule
&&$4$&&$13.735~043$&&$8.813~825$&&$8.795~449$&&$8.795~440$&&$8.795~440$&&$8.795~
440$&\cr\tablerule
&&$5$&&$11.686~537$&&$9.163~174$&&$9.163~095$&&$9.163~094$&&$9.163~093$&&$9.163~
093$&\cr\tablerule
&&$6$&&$11.110~897$&&$9.650~211$&&$9.646~713$&&$9.646~702$&&$9.646~701$&&$9.646~
701$&\cr\tablerule
&&$7$&&$11.145~653$&&$10.233~096$&&$10.225~061$&&$10.225~046$&&$10.225~045$&&$10
.225~045$&\cr\tablerule
&&$8$&&$11.492~447$&&$10.889~178$&&$10.879~092$&&$10.879~078$&&$10.879~077$&&$10
.879~077$&\cr\tablerule
&&$9$&&$12.020~404$&&$11.603~187$&&$11.592~993$&&$11.592~982$&&$11.592~982$&&$11
.592~982$&\cr\tablerule
&&$10$&&$12.662~990$&&$12.363~513$&&$12.354~191$&&$12.354~183$&&$12.354~183$&&$1
2.354~183$&\cr\tablerule
}}}}
\vfil\eject
% --------------------------------------------------------------------------

\noindent {\bf Table(III)}~~~Upper bounds $E_{21}^N$ for $A = 0$ and 
$H=-\Delta+x^2+{10\over x^{2.1}}$ for dimension $N=2$ to 10,
obtained by diagonalization of the $D\times D$ matrix $\cal{H}$
for $D=30$. The ``exact'' values $E_{21}$ were obtained by
direct numerical integration of Schr\"odinger's equation.
\bigskip

\hskip 1 true in
\vbox{\tabskip=0pt\offinterlineskip
\def\tablerule{\noalign{\hrule}}
\def\vr{\vrule height 12pt}
\halign to200pt{\strut#\vr&#
\tabskip=1em plus2em
&\hfil#\hfil
&\vrule#
&\hfil#\hfil
&\vrule#
&\hfil#\hfil
&\vr#\tabskip=0pt\cr
\tablerule&&$N$&&\bf $E_{21}^N$&&\bf $E_{21}$ &\cr\tablerule
&&2&&$16.543~648$&&$16.543~629$&\cr\tablerule
&&3&&$16.904~446$&&$16.904~445$&\cr\tablerule
&&4&&$17.381~709$&&$17.381~708$&\cr\tablerule
&&5&&$17.955~446$&&$17.955~444$&\cr\tablerule
&&6&&$18.607~070$&&$18.607~067$&\cr\tablerule
&&7&&$19.320~693$&&$19.320~691$&\cr\tablerule
&&8&&$20.083~407$&&$20.083~406$&\cr\tablerule
&&9&&$20.885~022$&&$20.885~021$&\cr\tablerule
&&10&&$21.717~608$&&$21.717~608$&\cr\tablerule
}}
% -------------------------------------------------------------------
\hfil\vfil\break
\vfil\eject
{\noindent Table(IV): Eigenvalues of Schr\"odinger's equation
$H\psi=E\psi$, where $H=-{{d^2}\over {dx^2}}+x^2+{\lambda\over x^{2.5}}$
obtained by diagonalization of the $n\times n$ matrix $\cal{H}$,
$n=1,2,\dots,5$ and minimizing over the parameter $A$.

\bigskip
\centerline{
\vbox{\tabskip=0pt\offinterlineskip
\def\tablerule{\noalign{\hrule}}
\def\vr{\vrule height 12pt}
\halign to360pt{\strut#\vr&#
\tabskip=0em plus1em
&\hfil#\hfil&\vrule#
&\hfil#\hfil&\vrule#
&\hfil#\hfil&\vrule#
&\hfil#\hfil&\vrule#
&\hfil#\hfil&\vrule#
&\hfil#\hfil&\vrule#
&\hfil#\hfil&\vrule#\tabskip=0pt\cr
\tablerule&&$\lambda$&&$E^{(1)}$&&$E^{(2)}$&&$E^{(3)}$&&$E^{(4)}$&&$E^{(5)}$&
&$E$&
\cr\tablerule
&&$1000$&&$44.959~423$&&$44.955~722$&&
$44.955~562$&&$44.955~559$&&$44.955~517$&&$44.955~485$&\cr\tablerule
&&$100$&&$17.546~305$&&$17.542~630$&&$17.542~082$&&$17.542~066$&&$17.542~040$&&
$17.541~890$
&\cr\tablerule
&&$10$&&$7.40~873$&&$7.736~864$&&$7.735~869$&&$7.735~645$&&$7.735~596$&&$7.735
~111
$&\cr\tablerule
&&$5$&&$6.302~942$&&$6.298~821$&&$6.297~638$&&$6.297~281$&&$6.297~145$&&$6.296~4
73
$&\cr\tablerule
&&$1$&&$3.325~682$&&$4.321~615$&&$4.320~076$&&$4.319~376$&&$4.318~963$&&$4.317~3
12
$&\cr\tablerule
&&$0.5$&&$3.857~330$&&$3.853~611$&&$3.852~085$&&$3.851~313$&&$3.850~823$&&$3.848
~553
$&\cr\tablerule
&&$0.1$&&$3.273~542$&&$3.271~566$&&$3.270~632$&&$3.270~082$&&$3.269~700$&&$3.266
~871
$&\cr\tablerule
&&$0.05$&&$3.157~126$&&$3.155~956$&&$3.155~378$&&$3.155~022$&&$3.154~768$&&$3.15
2~420
$&\cr\tablerule
&&$0.01$&&$3.037~845$&&$3.037~674$&&$3.037~581$&&$3.037520$&&$3.037~474$&&$3.036
~665
$&\cr\tablerule
&&$0.005$&&$3.019~610$&&$3.019~553$&&$3.019~522$&&$3.019~500$&&$3.019~484$&&$3.0
19~086
$&\cr\tablerule
&&$0.001$&&$3.004~053$&&$3.004~050$&&$3.004~049$&&$3.004~048$&&$3.004~047$&&$3.0
04~014
$&\cr\tablerule}}
}}

\vfil\eject

\hbox{\vbox{\psfig{figure=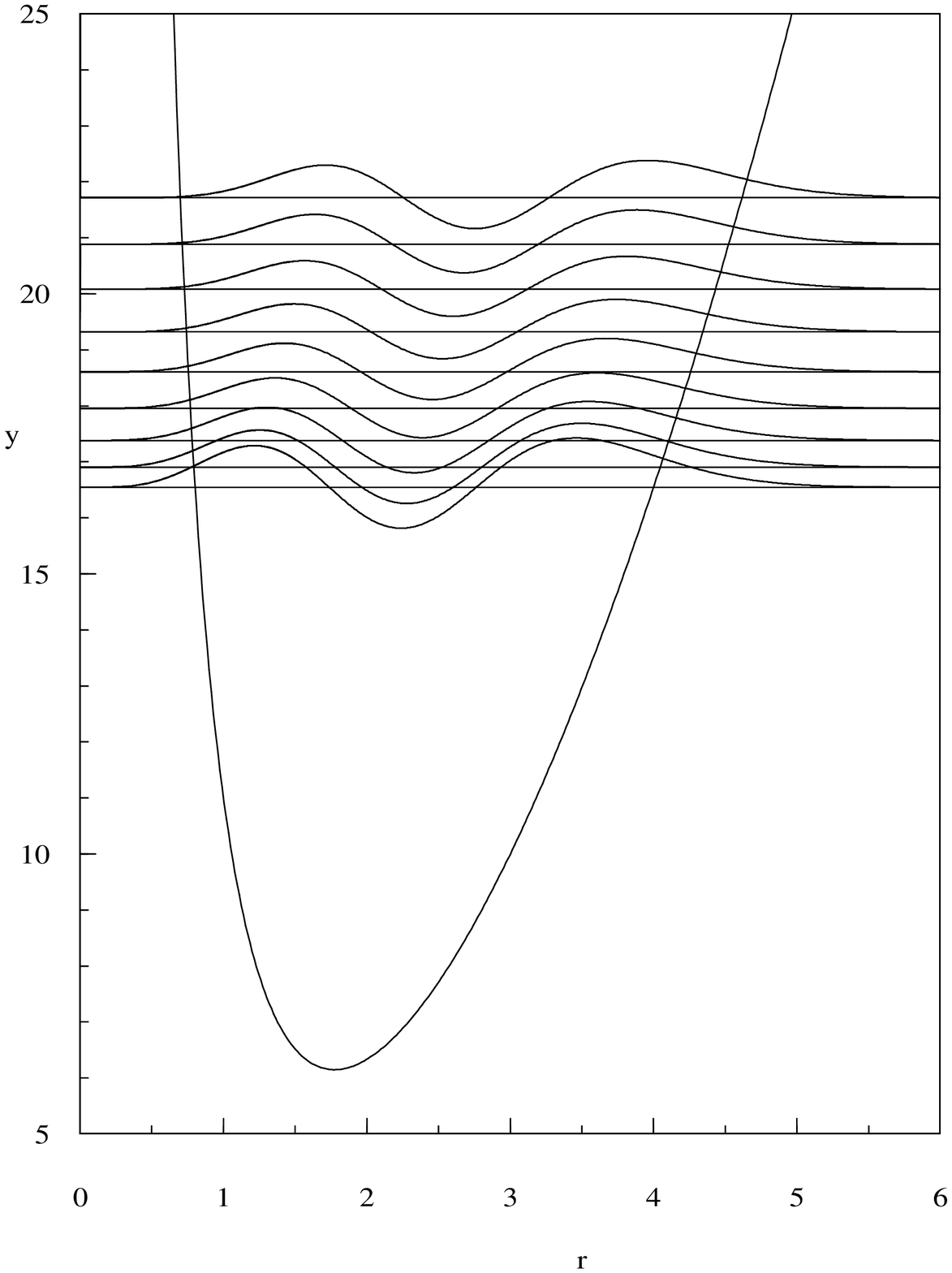,height=6in,width=5in,silent=}}}
%\pdfimage width 5in height 6in {nifig5.pdf}

\noindent {\bf Figure~(1)}~~The potential $y = V(r),$ and the nine wave
functions corresponding to the eigenvalues $E_{n\ell} = E_{21}$ for
spatial dimensions $N=2$ to $10.$

\hfil\vfil
\end